# Metal halide perovskite toxicity effects on plants are caused by iodide ions

Eline M. Hutter[1,2#], Reiny Sangster[3#], Christa Testerink[3], Bruno Ehrler[1*], Charlotte M. M. Gommers[3*]

1. Center for Nanophotonics, AMOLF, Science Park 104, 1098 XG Amsterdam, the Netherlands
2. Department of Chemistry, Utrecht University, Princetonlaan 6, 3584 CB Utrecht, the Netherlands
3. Laboratory of Plant Physiology, Wageningen University & Research, Droevendaalsesteeg 1, 6708 PB Wageningen, the Netherlands

\# These authors contributed equally

\* charlotte.gommers@wur.nl, B.Ehrler@amolf.nl

**Abstract**

Highly-efficient solar cells containing lead halide perovskites are expected to revolutionize sustainable energy production in the coming years. Combining these next-generation solar panels with agriculture, can optimize land-use, but brings new risks in case of leakage into the soil. Perovskites are generally assumed to be toxic because of the lead (Pb), but experimental evidence to support this prediction is scarce. We used *Arabidopsis thaliana* to test the toxicity of the lead-based perovskite $MAPbI_3$ (MA = $CH_3NH_3$) and several of its precursors in plants. Our results show that $MAPbI_3$ severely hampers plant growth at concentrations above 5 µM. Surprisingly, we find that the precursors MAI is equally toxic, while lead-based precursors without iodide are only toxic above 500 µM. These observations reveal that perovskite toxicity at low concentrations is caused by iodide ions specifically, and contrast the widespread idea that lead is the most harmful component. We calculate that iodide toxicity thresholds are likely to reach in the soil upon perovskite leakage, but much less so for lead toxicity thresholds. Hence, this work stresses the importance to further understand and predict harmful effects of iodide-containing perovskites in the environment.

**Introduction**

Sunlight provides enough energy to fulfill the global demand, making solar cells the most promising route towards a sustainable economy. Combing solar panels with agriculture, named *agri(photo)voltaics*, maximizes land use, while making optimum use of the sunlight for both crops and power generation (Figure 1A).[1] Solar cells based on both silicon and lead halide perovskites (LHPs) are considered the next-generation commercial solar cells, with potential efficiencies up to >45%.[2,3] However, placing such next-generation solar cells on *e.g.* agricultural lands raises questions about the safety of LHPs for the environment. That is, if unintentional leakage releases $Pb^{2+}$ ions in the soil, this may be harmful either to plants themselves or to humans and livestock through consumption of contaminated crops.[4] A few studies have specifically tested the toxicity of the LHP $MAPbI_3$ (MA = $CH_3NH_3$) and its supposedly less toxic tin-based counterpart on both plants and animals.[5–8] These studies focused on the heavy metals, but the presence of halides also raises environmental concerns,[9,10] which has not been studied in plants to date.

In this work, we find that the iodide in $MAPbI_3$ causes greater harm to plants than the lead. Model species *Arabidopsis thaliana* was used to study the effect of $MAPbI_3$ and different perovskite precursors on the growth and development of plants. Unlike previous work on perovskite toxicity,[5,6] we avoid acidification effects by buffering the growth media to maintain a fixed pH, so we could selectively study the effect of the perovskite and its precursors. We used controlled growth settings with a high-resolution range of salt concentrations to define the exact toxicity thresholds. We find that the presence of $MAPbI_3$ in the growth medium starts to affect plant performance at 5 µM and becomes significant at 10 µM. This concentration (10 µM) is exceeded, according to our calculations, for a practical situation where a perovskite solar cell leaks 10 cm deep into the soil of the same area. In contrast to previous works, we conclude from experiments comparing the precursors MAI and $PbI_2$ to MABr and $PbBr_2$, that the iodide is responsible for inhibiting *Arabidopsis* development, before toxicity effects of lead appear. These results show that a more rigorous assessment on the potential harmfulness of LHPs is needed and stress the importance of developing strategies to avoid halides from being released into the environment.

**Results and discussion**

Perovskites are fabricated from a lead halide salt, such as $PbI_2$, and an organic halide salt, such as MAI (Figure 1A): MAI + $PbI_2$ → $MAPbI_3$

Vice versa, decomposition of $MAPbI_3$ leads to the formation of its precursors $PbI_2$ and MAI. In order to assess the toxicity of LHPs, we germinated seeds of *Arabidopsis* (ecotype Columbia-0) on media containing different concentrations of $MAPbI_3$, and several precursor salts. The media were buffered at a pH of 5.8, to avoid acidification effects.[6] While seed germination is not affected, both $MAPbI_3$ and $PbI_2$ significantly inhibit plant growth at the seedling stage (depicted as rosette diameter) at concentrations > 10 µM (Figure 1B-C). Growth inhibition stagnated from 50 µM, suggesting that no additional toxicity occurs beyond this concentration. To specify the level of lead perovskite toxicity, we grew plants at a range of MAPbI$_3$ concentrations around 10 µM. As shown in figure 1D, growth is significantly inhibited by concentrations > 5 µM. In addition, plants appeared bleached when treated with higher concentrations of MAPbI$_3$ or PbI$_2$ (Figure 1B), which is supported by reduced chlorophyll levels in plants treated with over 10 µM of MAPbI$_3$ (Figure 1E).

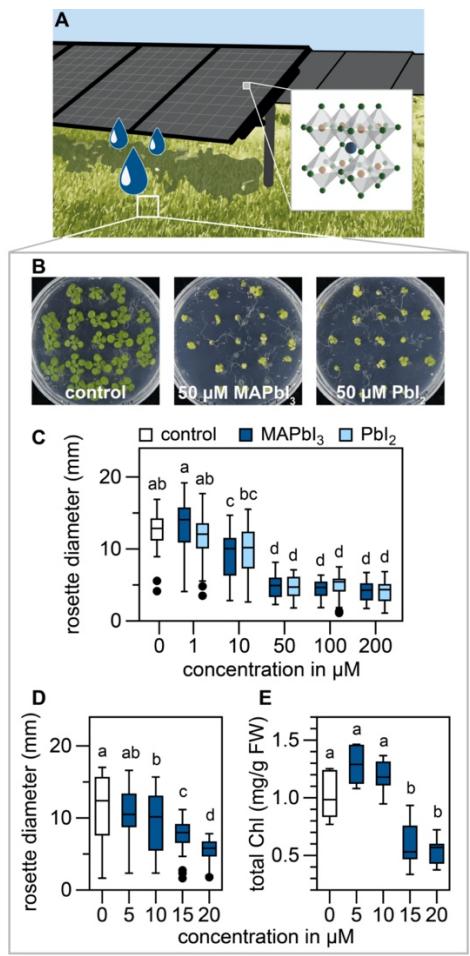

**Figure 1.** A) Schematic representation of agrivoltaics. The inset shows the structure of a lead halide perovskite. B) Representative pictures of 18 day-old *Arabidopsis* plants, grown on control growth medium, or supplemented with methylammonium lead iodide (MAPbI$_3$) or precursor PbI$_2$. C) Maximum rosette (plant) diameter in mm of 18 day-old *Arabidopsis* plants grown on growth medium supplemented with various concentrations of MAPbI$_3$ or PbI$_2$. D) Maximum rosette diameter in mm and E) total chlorophyll (ChlA + ChlB) content in mg/g fresh weight (FW) of 18 day-old *Arabidopsis* plants treated with various concentrations of MAPbI$_3$. For C – E: Boxes represent the median +/- 25%, bars +/- 50%, black dots are outliers. Different letters indicate statistical differences, with $p < 0.05$, tested by 2-way ANOVA and post-hoc Tukey test.

The $Pb^{2+}$ oxidation state is toxic to plants.[4,7,11] However, at concentrations for which we observed toxicity for both $MAPbI_3$ and $PbI_2$ (Figure 1C), lead nitrate ($Pb(NO_3)_2$) and another lead halide precursor ($PbBr_2$) did not affect plant growth (Figure 2A-B). We found these lead-containing salts to significantly impede *Arabidopsis* growth only at concentrations > 750 µM (Figure 2C). This effect could be directly attributed to the lead, as MABr did not affect plant growth at these concentrations. Interestingly, even though growth was inhibited at high concentrations, $Pb(NO_3)_2$ and $PbBr_2$ did not cause plant bleaching similar to $MAPbI_3$ and $PbI_2$ (compare Figure 1B and 2C). The observation that $MAPbI_3$ and $PbI_2$ hamper *Arabidopsis* growth at one order of magnitude lower concentrations (Figure 1) suggests that the iodide is at least in part responsible for this hampered growth. We repeated the growth experiments using the iodide precursor MAI and compared this to its bromide equivalent MABr (Figure 2D). As before, we find a significant inhibition of rosette size at 50 µM of MAI. In contrast, *Arabidopsis* growth is not affected by any concentration of MABr up to 1000 µM (Figure 2C and D).

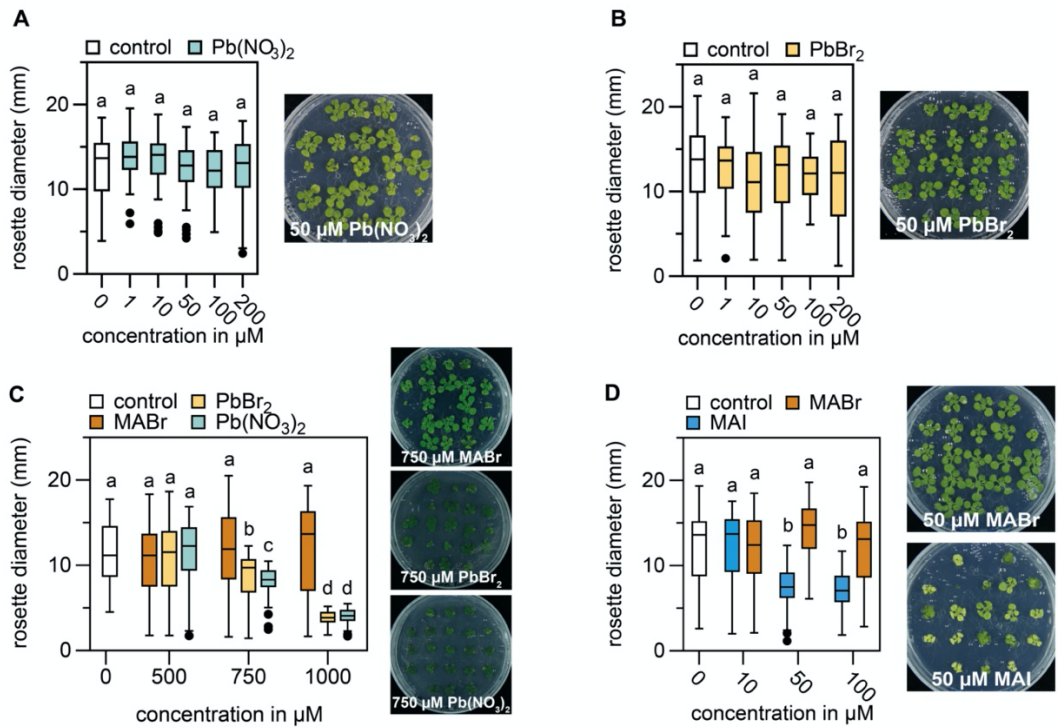

**Figure 2.** Maximum rosette (plant) diameter in mm of 18 day-old *Arabidopsis* plants grown on control growth medium, or supplemented with various concentrations of A) $Pb(NO_3)_2$ and B) $PbBr_2$, C) MABr, $Pb(NO_3)_2$ and $PbBr_2$, and D) MABr and MAI. In A – D), pictures show representative plants. Boxes represent the median +/- 25%, bars +/- 50%, black dots are outliers. Different letters indicate statistical differences, with $p < 0.05$, tested by 2-way ANOVA and post-hoc Tukey test.

Toxicity of MAPbI$_3$ starts at 10 μM, and between 10 μM and 50 μM for MAI (figure 2D), which we attribute to the three-fold higher concentration of available iodide ions in the MAPbI$_3$ treatment. The hypothesis that the iodide is responsible for the toxicity at low MAPbI$_3$ concentrations is further confirmed by the absence of toxicity of MABr and PbBr$_2$ in the same concentration ranges (Figure 2C). While iodide is considered a (micro-) nutrient,[12] the toxicity effects of iodide at higher concentrations are poorly understood. Our conclusion that that the low toxicity threshold of MAPbI$_3$ is caused by the presence of iodide, rather than lead, stresses the importance of further investigating the environmental effects of using iodide salts in solar panels.

Likely, the extent of toxicity of these halide salts in the environment will depend on the plant species, soil type, and depth to which the salts penetrate the soil in case of leakage. We estimate that a solar panel with a 400 nm thick perovskite layer contains 0.26 μmol of Pb$^{2+}$ and 0.79 μmol of iodide (I$^-$) per cm$^2$. If the full panel would leak and be homogeneously distributed over a water column of the same area and 10 cm deep, this will yield concentrations of 26 μM (<5 mg kg$^{-1}$ soil) Pb$^{2+}$ and 79 μM iodide. Hence, the expected maximum concentration of lead in case of leakage is much lower than the concentration (> 750 μM) at which it is toxic to the plant, and far below soil quality standards that range from 100 to 530 mg kg$^{-1}$ depending on the country.[5] The concentration of iodide (79 μM), on the other hand, exceeds the toxicity threshold of 50 μM, so that the release of iodide into the environment upon leakage of a perovskite solar panel may affect plant growth and fitness.

To conclude, although the perovskite community so far has mainly discussed the potential toxicity of lead,[5,7,8,13] we found that iodide is toxic to *Arabidopsis* plants at one order of magnitude lower concentrations. We further find that lead is only significantly toxic at levels far above those expected for a solar module failure. Our observations stress the importance of getting a more complete picture on the potential harmfulness of solar panels that contain LHPs, especially if these are placed on agricultural lands, and to develop strategies to prevent the release of halides into the environment. The latter should primarily be realized by careful encapsulation of the solar cells to prevent leakage, or in case of unforeseen calamities, by considering halide-tolerant plants that accumulate leaked halides in a phytoremediation approach to clean the soil.[10]

**Materials and Methods**

*Plant material and growth conditions*

*Arabidopsis thaliana* ecotype Columbia-0 was used for all experiments. Seeds were surface-sterilized (20% bleach, 0.5%SDS), rinsed with sterile water and sown on half-strength Murashige & Skoog medium including vitamins (Duchefa Biochemie), containing 0,1% MES monohydrate buffer (Duchefa Biochemie) and 1% v/w Daishin agar (Duchefa Biochemie). Salts where added from a 1 mM stock in the desired concentration before autoclaving. The pH of all media was set at 5.8 using 0.1 N KOH to prevent any harmful acidification effects as described by Babayigit et al.[6] Seeds were sown on petri dishes (9 cm diameter) containing 20 mL medium, stratified (4 ºC, dark) for 4 days to synchronize germination, and afterwards placed in a climate chamber (16 hours light period, 22 ºC, photosynthetic active radiation 150 µmol photons/m$^2$/s).

*Perovskite solutions*

$MAPbI_3$ was prepared by mixing the precursor powders in a stoichiometric ratio (0.4 millimole of MAI and 0.4 millimole of $PbI_2$) until a black powder was obtained. Consequently, deionized water was added to the powder. On the addition of water, the powder immediately turned yellow, indicating decomposition of the perovskite into MAI (high solubility in water) and $PbI_2$ (yellow, poor solubility in water). In addition, the separate precursors $PbI_2$, $PbBr_2$, MAI and MABr were dissolved in water at concentrations of 1 mM, as well as $Pb(NO_3)_2$ for reference experiments. The solutions for the experiments were prepared by dilution of the above stock solution.

*Plant growth assays and chlorophyll quantification*

Pictures of the plants were taken after 18 days of growth (2 petri dishes per treatment, 21 plants per petri dish). Maximum rosette diameter of all plants was analyzed using FIJI software. From each petri dish, three plants were harvested, weighed and used for chlorophyll quantification: plants were shaken in 1 mL DMSO at 65 °C in darkness for 1 hour, cooled and kept at room temperature for 30 minutes. Absorption was measured at 664 and 647 nm using a SpectraMax® Plus 384 Microplate Reader (Molecular Devices). Concentration of chlorophyll A and B was calculated by: [ChlA] (mg/L) = 12.25*(A664) – 2.55*(A647); [ChlB] (mg/L) = 20.31*(A647) – 4.91*(A664).[14] Values were converted into mg per mg fresh weight.

*Statistics*

Statistical analyses were performed using the MVApp.[15] Levene's test was used to test for equal variances. Multivariate comparisons were made with ANOVA, followed by a post-hoc Tukey test.

*Estimated concentrations in the soil*

Assuming a density of 4.09 gcm$^{-3}$ for $MAPbI_3$,[5] and a thickness of 400 nm (4x10$^{-5}$ cm), a solar panel contains 1.6x10$^{-4}$ g perovskite per cm$^2$. With molar weight fractions of 0.05 for MA$^+$, 0.33 for Pb$^{2+}$ and 0.61 for I$^-$, this translates to 8.5x10$^{-6}$ g MA$^+$, 5.47x10$^{-5}$ g Pb$^{2+}$ and 1 x 10$^{-4}$ g I$^-$ per cm$^2$ solar panel. If a full solar panel would break and leak into an equivalent area, the concentration in the soil will depend on the depth of leakage. For example, homogeneous distribution of the perovskite over the first 10 cm soil leads to concentrations of 8.5x10$^{-7}$ g/cm$^3$ MA$^+$, 5.47x10$^{-6}$ g/cm$^3$ Pb$^{2+}$ and 1 x 10$^{-5}$ g/cm$^3$ I$^-$. Dividing these numbers by the respective molar weights, and considering that 1 cm$^3$ = 1 mL, yields molarities of 26 µM MA$^+$, 26 µM Pb$^{2+}$ and 79 µM

I$^-$. If the soil has a density between 1 and 2 g/cm$^3$, the concentrations are between 5 and 2.7 mgkg$^{-1}$ for Pb$^{2+}$ and between 10 and 5 mgkg$^{-1}$ for the I$^-$.

**Acknowledgments**

We thank Mariëlle Schreuder for technical assistance. The Dutch Research Council (NWO) is acknowledged for funding.

**Author Contributions**

E.M.H, C.T., B.E. and C.M.M.G. conceived the project; E.M.H. and R.S. performed the experiments; E.M.H, R.S. and C.M.M.G. analyzed the data; E.M.H, C.T., B.E. and C.M.M.G. wrote the manuscript.

**References**

(1) Adeh, E. H.; Good, S. P.; Calaf, M.; Higgins, C. W. Solar PV Power Potential Is Greatest Over Croplands. *Sci. Rep.* **2019**, *9* (1), 1–6. https://doi.org/10.1038/s41598-019-47803-3.

(2) Futscher, M. H.; Ehrler, B. Efficiency Limit of Perovskite/Si Tandem Solar Cells. *ACS Energy Lett.* **2016**, *1*, 863–868. https://doi.org/10.1021/acsenergylett.6b00405.

(3) Leijtens, T.; Bush, K. A.; Prasanna, R.; McGehee, M. D. Opportunities and Challenges for Tandem Solar Cells Using Metal Halide Perovskite Semiconductors. *Nat. Energy* **2018**, *3* (10), 828–838. https://doi.org/10.1038/s41560-018-0190-4.

(4) Pourrut, B.; Shahid, M.; Dumat, C.; Winterton, P.; Pinelli, E. Lead Uptake, Toxicity, and Detoxification in Plants. *Rev. Environ. Contam. Toxicol.* **2011**, *213*. https://doi.org/10.1007/978-3-319-10479-9.

(5) Li, J.; Cao, H. L.; Jiao, W. Bin; Wang, Q.; Wei, M.; Cantone, I.; Lü, J.; Abate, A. Biological Impact of Lead from Halide Perovskites Reveals the Risk of Introducing a Safe Threshold. *Nat. Commun.* **2020**, *11* (1), 1–5. https://doi.org/10.1038/s41467-019-13910-y.

(6) Babayigit, A.; Duy Thanh, D.; Ethirajan, A.; Manca, J.; Muller, M.; Boyen, H. G.; Conings, B. Assessing the Toxicity of Pb-and Sn-Based Perovskite Solar Cells in Model Organism Danio Rerio. *Sci. Rep.* **2016**, *6*, 1–11. https://doi.org/10.1038/srep18721.

(7) Babayigit, A.; Ethirajan, A.; Muller, M.; Conings, B. Toxicity of Organometal Halide Perovskite Solar Cells. *Nat. Mater.* **2016**, *15* (3), 247–251.

(8) Hailegnaw, B.; Kirmayer, S.; Edri, E.; Hodes, G.; Cahen, D. Rain on Methylammonium Lead Iodide Based Perovskites: Possible Environmental Effects of Perovskite Solar Cells. *J. Phys. Chem. Lett.* **2015**, *6* (9), 1543–1547. https://doi.org/10.1021/acs.jpclett.5b00504.

(9) Medrano-Macías, J.; Leija-Martínez, P.; González-Morales, S.; Juárez-Maldonado, A.; Benavides-Mendoza, A. Use of Iodine to Biofortify and Promote Growth and Stress Tolerance in Crops. *Front. Plant Sci.* **2016**, *7* (AUG2016), 1–20. https://doi.org/10.3389/fpls.2016.01146.

(10) Incrocci, L.; Carmassi, G.; Maggini, R.; Poli, C.; Saidov, D.; Tamburini, C.; Kiferle, C.; Perata, P.; Pardossi, A. Iodine Accumulation and Tolerance in Sweet Basil (Ocimum Basilicum L.) With Green or Purple Leaves Grown in Floating System Technique. *Front. Plant Sci.* **2019**, *10* (December), 1–15. https://doi.org/10.3389/fpls.2019.01494.

(11)     Gupta, D. K.; Chatterjee, S.; Walther, C. *Lead in Plants and the Environment*; 2020; Vol. 5. https://doi.org/10.1007/978-3-030-21638-2.

(12)     Kiferle, C.; Martinelli, M.; Salzano, A. M.; Gonzali, S.; Salvadori, P. A.; Hora, K.; Holwerda, H. T. Evidences for a Nutritional Role of Iodine in Plants. *bioRxiv Prepr.* **2020**. https://doi.org/10.1101/2020.09.16.300079.

(13)     Slavney, A. H.; Smaha, R. W.; Smith, I. C.; Jaffe, A.; Umeyama, D.; Karunadasa, H. I. Chemical Approaches to Addressing the Instability and Toxicity of Lead–Halide Perovskite Absorbers. *Inorg. Chem.* **2016**, *56* (1), 46–55. https://doi.org/10.1021/acs.inorgchem.6b01336.

(14)     Porra, R. J.; Thompson, W. A.; Kriedemann, P. E. Determination of Accurate Extinction Coefficients and Simultaneous Equations for Assaying Chlorophylls a and b Extracted with Four Different Solvents : Verification of the Concentration of Chlorophyll Standards by Atomic Absorption Spectroscopy. *Biochim. Biophys. Acta* **1989**, *975*, 384–394.

(15)     Julkowska, M. M.; Saade, S.; Agarwal, G.; Gao, G.; Pailles, Y.; Morton, M.; Awlia, M.; Tester, M. MVApp — Multivariate Analysis Application for Streamlined Data Analysis and Curation. *Plant Physiol.* **2019**, *180*, 1261–1276. https://doi.org/10.1104/pp.19.00235.